\documentclass[twoside,fleqn]{article}
\usepackage{espcrc2}
\usepackage{times,mathptm}
\usepackage[dvips]{graphicx}
\newcommand{\fcaption}[1]{%
\vspace*{-1.1cm}
\caption{#1}
\vspace*{-0.1cm}
}
\title{Center Dominance Recovered: Direct Laplacian Center Gauge%
\thanks{Presented by \v{S}.\ Olej\-n{\'\i}k.  Supported in part by 
the NATO Collaborative Linkage Grant No.\ PST.CLG.976987
and by the Slovak Grant Agency for Science (Grant VEGA No.\ 2/7119/2000).}}
\author{%
Manfried Faber\address{Atominstitut der \"osterreichischen Universit\"aten,
Tech.\ Univ.\ Wien, A--1040 Vienna, Austria}, 
Jeff Greensite\address{Physics and Astronomy Dept., San Francisco State Univ.,
San Francisco, CA~94117, USA}
and 
\v{S}tefan Olej\-n{\'\i}k\address{Institute of Physics, Slovak Academy 
of Sciences, SK--842 28 Bratislava, Slovakia}}
\begin{document}
%
% ABSTRACT
%
\begin{abstract}
  We introduce a variation of direct maximal center gauge
  fixing: the ``direct Laplacian'' center gauge.  The new procedure 
  consists of first fixing to the Laplacian adjoint Landau gauge, followed by
  overrelaxation to the nearby Gribov copy of the direct maximal center gauge.
  Certain shortcomings of maximal center gauge,
  associated with Gribov copies, are overcome in the new gauge, in
  particular center dominance is recovered.
\end{abstract}
\maketitle
%
% SECTION 1
%
\section{DRAMA OF GRIBOV COPIES: THE FIRST FIVE ACTS}\label{drama}
	We proposed a procedure for identifying center vortices
in lattice configurations based on center projection in maximal 
center (or adjoint Landau) gauge, and accumulated evidence in favor 
of the  center vortex model of color 
confinement~\cite{DelDebbio:1998uu}. 
MCG fixing, however, suffers from the Gribov copy problem: The 
iterative gauge fixing procedure converges to a local maximum 
which is slightly different for every gauge copy of a given 
lattice configuration. The problem seemed quite innocuous at first; 
we observed that vortex locations in 
random copies of a given configuration are strongly correlated.
Recently, the successes of the approach based on MCG fixing 
have been overshadowed by serious difficulties, so serious
that they led the authors of Ref.~\cite{Bornyakov:2001ig} to speak about 
the \textit{Gribov copy drama}. Indeed, it follows the structure of the 
classical ancient tragedy:

1.\ \textit{Exposition:}
The projected string tension in MCG was shown to reproduce the full asymptotic 
string tension (\textit{center dominance})
\cite{DelDebbio:1998uu}.

2.\ \textit{Complication:}
Kov\'acs and Tom\-bou\-lis~\cite{Kovacs:1999st} 
observed that if we fix to 
\textit{Landau gauge}, before overrelaxation to MCG, 
center do\-mi\-nan\-ce is lost!

3.\ \textit{Climax:}
Bornyakov et al.~\cite{Bornyakov:2000cd} pointed out strong dependence
of results on the number of gauge copies used in MCG maximization. They chose
the best from $N_{cop}$ copies and extrapolated for {$N_{cop}\to\infty$}:
the full string tension was underestimated by about {30\%}!

4.\ \textit{Reversal:}
It was shown~\cite{Bertle:2000qv} 
that the results depend on the number of gauge copies
\textit{and the lattice size}. On a large lattice 
(compared to the typical size of the vortex core) the problem 
disappears.

5.\ \textit{Catastrophe:}
Bornyakov et al.~\cite{Bornyakov:2001ig} 
observed that using \textit{simulated annealing} one can find better
maxima than with the usual method of \textit{overrelaxation}, 
but the center projected
string tension again is only about 2/3 of the full string tension. 

	Our aim in this paper is to argue that, unlike ancient tragedies,
the Gribov copy drama can have the sixth act, a happy ending. We 
shall introduce a new procedure, direct Laplacian center gauge (DLCG), 
that overcomes the difficulties associated with Gribov copies. Before doing 
that, however, we shall briefly review a new insight into MCG fixing,
due to~\cite{Engelhardt:2000xw,Faber:2001hq}.
%
% SECTION 2
%
\section{GAUGE FIXING AS FINDING A ``BEST FIT''}\label{fit}
	Imagine running a Monte Carlo simulation and asking for the
pure gauge configuration closest, in configuration space, to the
given thermalized lattice. It is easy to show that finding such an ``optimal'' 
configuration is equivalent to fixing to the \textit{Landau gauge}.

	Allow now for {$Z_2$} dislocations 
in the gauge transformation, i.e.\ fit the lattice configuration
by one with thin center vortices:
\begin{equation}\label{vortex_conf}
U_\mu^{vor}(x)\ \equiv\  
g(x) Z_\mu(x)g^\dagger(x+\hat{\mu}), 
\quad Z_\mu(x)=\pm 1.
\end{equation}
Such $U_\mu^{vor}(x)$ is a continuous pure gauge in the adjoint 
representation, which is blind to the $Z_\mu(x)$ factor. Then 
the fit can be easily performed in two steps:

	1. one can determine $g(x)$ up to a $Z_2$ 
transformation by minimizing the square distance between
$U_{A\mu}$ and $U_{A\mu}^{vor}$ in the adjoint representation -- 
this is equivalent to fixing to the adjoint Landau gauge, aka \textit{direct 
maximal center gauge};

	2. $Z_\mu(x)$ is then fixed by the 
\textit{center projection prescription}:
$Z_\mu(x)=\mbox{sign}\;\mbox{Tr}
\left[g^\dagger(x) U_\mu(x) g(x+\hat{\mu})\right]$.

	It is now immediately clear that $U_\mu^{vor}(x)$ is a bad
fit to $U_\mu(x)$ at links belonging to thin vortices (i.e.\ to the
P-plaquettes formed from $Z_\mu(x)$). We recall that a 
plaquette $p$ is a P-plaquette iff $Z(p)=-1$ (where $Z(C)$ denotes the
product of $Z_\mu(x)$ around the contour~$C$) and that  P-plaquettes belong
to P-vortices. Let us write the gauge transformed configuration as 
\begin{equation}
{}^g U_\mu(x) = Z_\mu(x)\;e^{iA_\mu(x)},
\qquad \mbox{Tr}\;e^{iA_\mu(x)}\ge0.
\end{equation}
At large $\beta$ values
$\textstyle{\frac{1}{2}}\mbox{Tr}[U_P]=1-O(\frac{1}{\beta})$,
and equals to
\begin{equation}
\left(Z_P\right)
\textstyle{\frac{1}{2}}
\mbox{Tr}\displaystyle{\prod_P} e^{iA_\mu(x)} 
{= \atop \mbox{{\small on P-plaq.}}}
{(-1)}\times\textstyle{\frac{1}{2}}
\mbox{Tr}\displaystyle{\prod_P} e^{iA_\mu(x)}.
\end{equation}
The last equation implies that at least at one link belonging to the 
P-plaquette $A_\mu(x)$ cannot be small, therefore ${}^g U_\mu(x)$ 
must strongly deviate 
from the center element. This indicates that the quest for the global maximum 
may not be the best strategy; one should rather try to exclude contributions
from P-plaquettes where the fit is inevitably bad~\cite{Faber:2001hq}, 
or modify the gauge fixing procedure to soften the fit at vortex cores.
%
% SECTION 3
%
\section{DIRECT LAPLACIAN CENTER GAUGE}\label{dlcg}
	Our new proposal to overcome the Gribov copy problem
was inspired by the Laplacian Landau gau\-ge~\cite{Vink:1992ys}.
The idea is the following (for details see \cite{Faber:2001zs}):

	To find the ``best fit'' to a lattice configuration by a thin 
center vortex configuration one looks for a matrix $M(x)$ maximizing 
the expression:
\begin{equation}
{\cal{R}}_M= \displaystyle{\sum_{x,\mu}} 
\mbox{Tr}\left[M^T(x) U_{A\mu}(x) M(x+\hat{\mu})\right],
\end{equation}
with a constraint that $M(x)$ should be an SO(3) matrix in
any site $x$. We soften the orthogonality constraint by demanding it
only \textit{``on average''}:
%
%\begin{equation}
$\langle M^T\cdot M\;\rangle\equiv
\textstyle{\frac{1}{\cal V}\sum_x} M^T(x)\cdot M(x)= \mathbf{1}$.
%\end{equation}
%
	
	It is convenient to write the columns of $M(x)$ as a set of 
3-vectors $\vec{f}_a(x)$: $f^{b}_a(x) = M_{ab}(x)$.
The optimal $M(x)$ is determined by \textit{three lowest 
eigenvectors} $\vec{f}_j(x)$ of the covariant adjoint Laplacian operator:
\begin{equation}
\nonumber
{\cal{D}}_{ij}(x,y)= 
2D\delta_{xy}\delta_{ij}
- \displaystyle{\sum_{\pm\mu}}
\left[U_{A,\pm\mu}(x)\right]_{ij}\delta_{y,x\pm\hat\mu}.
\end{equation}

	The resulting real matrix field $M(x)$ has further to be
mapped onto an SO(3)-valued field $g_{A}(x)$. A \textit{\textbf{naive map}}
(which could also be called \textit{Laplacian adjoint Landau gauge})
amounts to choosing $g_{A}(x)$ closest to $M(x)$. Such
a map is well known in matrix theory and is called 
\textit{polar decomposition}.

	A better procedure, in our opinion, is the
\textit{\textbf{Laplacian map}}, that leads to 
\textit{direct Laplacian center gauge}. We try to locate $g_A(x)$ as 
close to $M(x)$ \textit{local} maximum of the MCG (constrained)
maximization problem. To achieve this, we first make the naive map
(polar decomposition),
then use the usual quenched maximization (overrelaxation) to relax to
the nearest (or at least nearby) maximum of the MCG fixing condition.
%
% SECTION 4
%
\section{CENTER DOMINANCE}\label{results}
	To test the procedure of the last section, we have recalculated the
vortex observables introduced in our previous work \cite{DelDebbio:1998uu}, 
with P-vortices located via center projection after
fixing the lattice to the new direct Laplacian center gauge. The full set of 
results has been published in \cite{Faber:2001zs}. Here we just concentrate
on the issue of center dominance which is crucial for the whole 
picture. Without being able to reproduce the string tension of the full
theory, one cannot claim to have isolated degrees of freedom that
are associated with the confinement mechanism.

\begin{figure}[!t]
\centering
{\scalebox{1.0}{\includegraphics[width=0.45\textwidth]{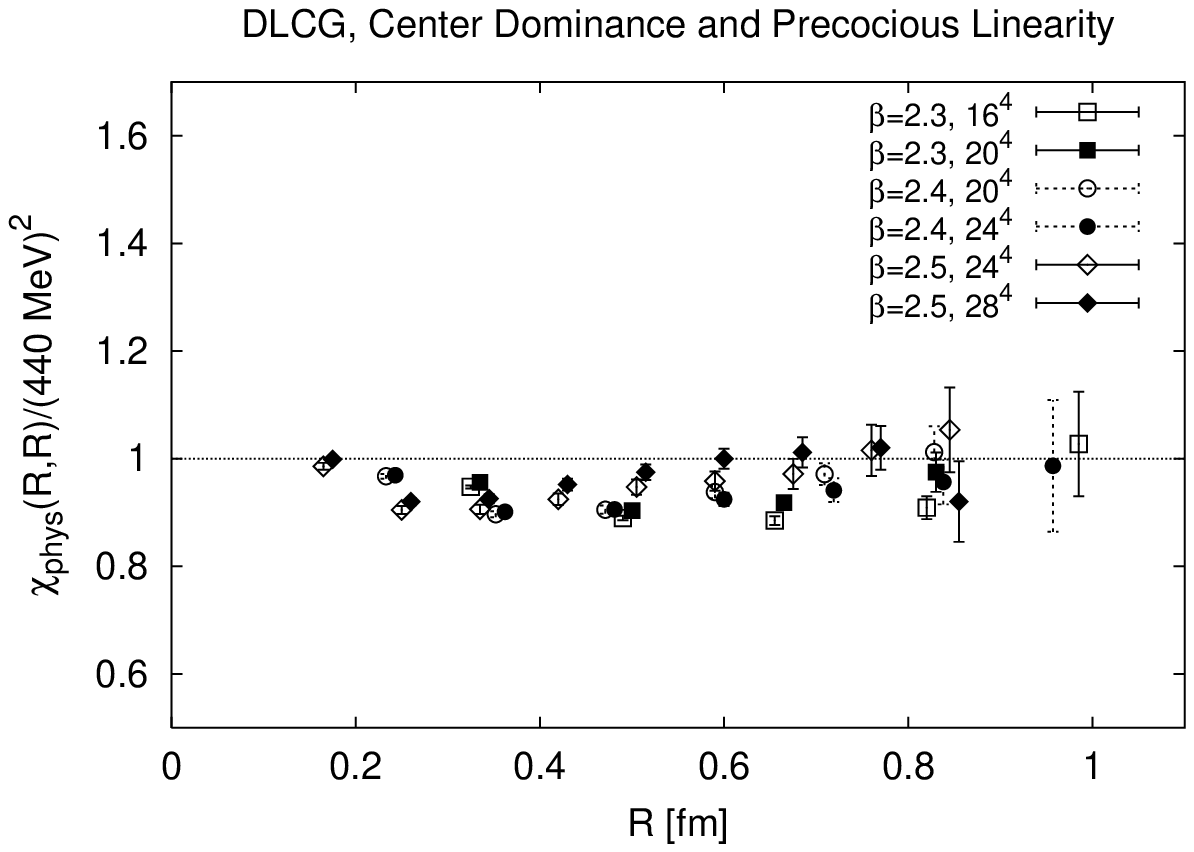}}}
\fcaption{The ratio of projected Creutz ratios to the full asymptotic
string tension, as a function of $R$.  
%The data is taken from $\chi_{cp}(R,R)$ at a variety of couplings and
%lattice sizes.
}\label{sigma2}
\end{figure}

	A subset of our results, obtained at a variety of couplings on
our largest lattice sizes, is shown in Fig.\ \ref{sigma2}. It is clearly 
seen that center dominance is restored in DLCG. Moreover, we once
again observe precocious linearity (the very weak
dependence of projected Creutz ratios on the distance $R$) signalling 
that the center projected degrees of freedom have isolated the long range
physics, and are not mixed up with ultraviolet fluctuations. Creutz 
ratios differ by at most about 10\% from the phenomenological value of the 
string tension.

	Also other physical results are reproduced in the new gauge:
scaling of the vortex density, agreement of ratios of vortex limited
Wilson loops with simple expectations, the fact that removing center vortices
from lattice configurations destroys confinement~\cite{Faber:2001zs}.
%
% SECTION 5
%
\section{COMPARISON TO INDIRECT LCG}
	To avoid problems with Gribov copies, de Forcrand et al.\ 
\cite{Alexandrou:2000iy} proposed the Laplacian center gauge. The 
procedure is similar to ours, but differs in a few 
important points. The gauge transformation matrix is found from 
\textit{two} lowest eigenvectors of the covariant adjoint Laplacian operator,
one first fixes SU(2) to the U(1) subgroup, 
and  center vortex surfaces should in principle be identified 
via ambiguities of the gauge fixing condition. However, there is
no good separation between confinement and short range physics in this 
gauge (center dominance seen only at largest distances, no precocious 
linearity, vortex density not scaling), and 
identification of vortices on a lattice
via gauge fixing ambiguities is practically impossible (center projection is 
necessary). Langfeld et al.\ \cite{Langfeld:2001nz} 
therefore used overrelaxation to MCG after fixing 
to LCG. This we would call \textit{indirect LCG} (ILCG).

\begin{figure}[t!]
\centering
{\scalebox{1.0}{\includegraphics[width=0.45\textwidth]{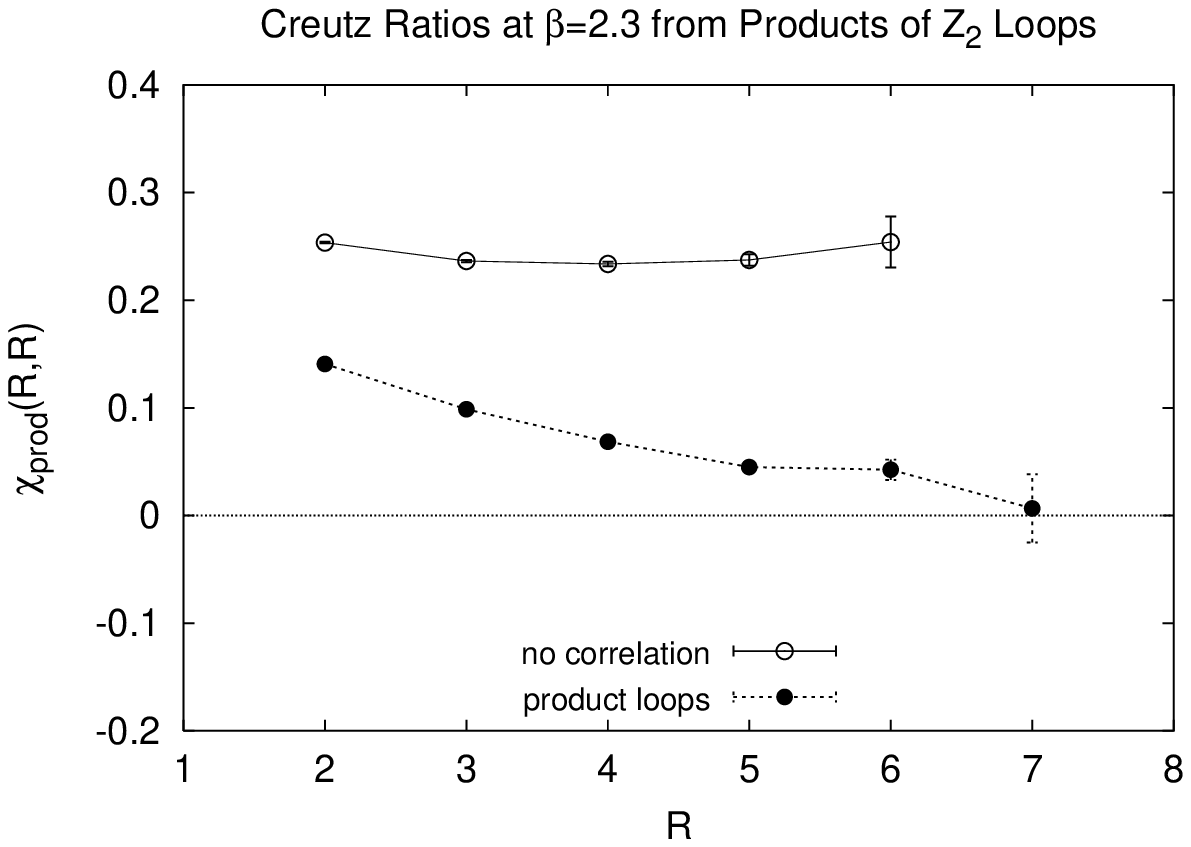}}}
\fcaption{Creutz ratios from ``product Wilson loops'' show strong correlation
between DLCG and ILCG.}\label{2p3corlcg}
\end{figure}

	It turns out that results from DLCG and ILCG are quite similar.
Center dominance, precocious linearity, and vortex density scaling are 
observed, however, Creutz ratios in ILCG are somewhat lower than those in 
DLCG. One can attribute the similarity of both procedures to the strong
correlation between P-vortex locations in both gauges. This is illustrated in 
Fig.\ \ref{2p3corlcg} showing Creutz ratios calculated from ``product 
Wilson loops'' $W_{prod}(C)=\langle Z_{ILCG}(C)\;Z_{DLCG}(C)\rangle$.
If there were no correlation, these ratios should approach twice the
asymptotic string tension, while they should go to zero for strong correlation.
Our data clearly favor the latter case.
\vfill
%
% SECTION 6
%
\section{CONCLUSION}\label{conclusion}
We have proposed a new gauge, DLCG, that combines
fixing to Laplacian adjoint Landau gauge with the usual 
quenched maximization. The first step of the procedure is unique, 
in the second step no strong gauge copy dependence appears. 
The procedure can be interpreted as a ``best fit'' of a lattice 
configuration by thin vortices,
softened at vortex cores.
Center dominance, precocious linearity and scaling of the vortex density
are recovered in the new gauge.
%
% REFERENCES
%

%

\end{document}